\documentclass[sigconf, natbib=false]{acmart}
\AtBeginDocument{%
  \providecommand\BibTeX{{%
    Bib\TeX}}}

\setcopyright{acmlicensed}
\copyrightyear{2018}
\acmYear{2018}
\acmDOI{XXXXXXX.XXXXXXX}
\acmConference[Conference acronym 'XX]{Make sure to enter the correct
  conference title from your rights confirmation email}{June 03--05,
  2018}{Woodstock, NY}
\acmISBN{978-1-4503-XXXX-X/2018/06}

\usepackage{cite}
\usepackage{algpseudocode}
\usepackage{algorithm}
\usepackage{graphicx}
\usepackage{textcomp}
\usepackage{xcolor}
\usepackage[most]{tcolorbox}
\usepackage{tikz}
\tcbuselibrary{skins, breakable}
\usetikzlibrary{arrows.meta, positioning}

\usepackage{comment}
\usepackage{lmodern}
\usepackage{parskip}

\graphicspath{{./figures/}}

\newcommand{\todo}[1]{{\color{red}TODO: #1}}

\newcommand{\savevspace}{\vspace{-.5em}}

\def\BibTeX{{\rm B\kern-.05em{\sc i\kern-.025em b}\kern-.08em
    T\kern-.1667em\lower.7ex\hbox{E}\kern-.125emX}}
\begin{document}

\title{Evaluating the Efficacy of LLM-Based Reasoning for Multiobjective HPC Job Scheduling 
}

\author{Prachi Jadhav}
\affiliation{%
  \institution{University of Tennessee, Knoxville\\Oak Ridge National Laboratory}
  \city{Oak Ridge, TN}
  \country{USA}}
\email{jadhavpr@ornl.gov}

\author{Hongwei Jin}
\affiliation{%
  \institution{Argonne National Laboratory}
  \city{Lemont, IL}
  \country{USA}}
\email{jinh@anl.gov}

\author{Ewa Deelman}
\affiliation{%
  \institution{University of Southern California}
  \city{Los Angeles, CA}
  \country{USA}}
\email{deelman@isi.edu}

\author{Prasanna Balaprakash}
\affiliation{%
  \institution{Oak Ridge National Laboratory}
  \city{Oak Ridge, TN}
  \country{USA}}
\email{pbalapra@ornl.gov}


\begin{abstract}
High-Performance Computing (HPC) job scheduling involves balancing conflicting objectives such as minimizing makespan, reducing wait times, optimizing resource use, and ensuring fairness. Traditional methods, including heuristic-based, e.g., First-Come-First-Served(FJFS) and Shortest Job First (SJF), or intensive optimization techniques, often lack adaptability to dynamic workloads and, more importantly, cannot simultaneously optimize multiple objectives in HPC systems. To address this, we propose a novel Large Language Model (LLM)-based scheduler using a ReAct-style framework (Reason + Act), enabling iterative, interpretable decision-making. The system incorporates a scratchpad memory to track scheduling history and refine decisions via natural language feedback, while a constraint enforcement module ensures feasibility and safety.
We evaluate our approach using OpenAI's O4-Mini and Anthropic's Claude 3.7 across seven real-world HPC workload scenarios, including heterogeneous mixes, bursty patterns, and adversarial cases etc. Comparisons against FCFS, SJF, and Google OR-Tools (on 10 to 100 jobs) reveal that LLM-based scheduling effectively balances multiple objectives while offering transparent reasoning through natural language traces. The method excels in constraint satisfaction and adapts to diverse workloads without domain-specific training. However, a trade-off between reasoning quality and computational overhead challenges real-time deployment.
This work presents the first comprehensive study of reasoning-capable LLMs for HPC scheduling, demonstrating their potential to handle multiobjective optimization while highlighting limitations in computational efficiency. The findings provide insights into leveraging advanced language models for complex scheduling problems in dynamic HPC environments.
\end{abstract}



\received{20 February 2007}
\received[revised]{12 March 2009}
\received[accepted]{5 June 2009}

\maketitle

\section{Introduction}
\label{sec:intro}
High-Performance Computing (HPC) systems enable scientific breakthroughs but require efficient job scheduling to be optimized. This NP-hard optimization problem~\cite{wang2021rlschert} demands rapid decisions balancing conflicting objectives: minimizing makespan, wait times, and fairness while maximizing utilization~\cite{abdurahman2024scalable,li2022mrsch}. Traditional approaches use heuristic algorithms (FCFS with backfilling)~\cite{srinivasan2002characterization} or optimization techniques, but struggle with dynamic workloads and hardware heterogeneity~\cite{goponenko2022metrics}. Deep Reinforcement Learning (DRL) shows promise but faces sample inefficiency and partial observability challenges~\cite{wang2021rlschert}.

Modern LLMs demonstrate remarkable reasoning capabilities through Chain-of-Thought (CoT) approaches~\cite{wang2025tutorial,anthropic2025hybrid}. We present the first LLM-based scheduler using a ReAct framework \cite{yao2023react} enhanced with a persistent scratchpad memory, that tracks decisions and feedback across timesteps, enabling interpretable multiobjective optimization without domain-specific training. Our hybrid architecture separates natural language reasoning from constraint enforcement, ensuring reliability while providing decision transparency critical for HPC operations.

\savevspace
\subsection{Related works}
HPC scheduling involves allocating heterogeneous jobs with resource demands (nodes, memory, walltime) to finite compute resources. The core challenge lies in optimizing makespan and utilization while handling continuous job arrivals~\cite{li2022mrsch}. Traditional approaches range from simple FCFS policies to multiobjective optimization strategies~\cite{srinivasan2002characterization, blazewicz2019handbook, zhao1989performance, simon2013multiple}. Moreover, classical metaheuristic methods like Genetic Algorithms (GA)~\cite{mirjalili2019genetic}, Simulated Annealing (SA)~\cite{bertsimas1993simulated}, and Particle Swarm Optimization (PSO)~\cite{wang2018particle} have been applied to HPC scheduling, primarily to optimize a single objective (e.g., makespan or throughput) through iterative search over job permutations ~\cite{blum2003metaheuristics}.

LLMs are emerging in HPC ecosystems for job script generation~\cite{yin2025chathpc} and policy design. Recent works explore LLM-based resource allocation frameworks~\cite{zhu2025llmsched} and multi-agent coordination~\cite{amayuelas2025self}. Key challenges include domain adaptation, computational costs, and constraint integration~\cite{chen2024landscape}. Moreover, the concept of LLM agents capable of planning and tool use suggests future applications in complex decision-making tasks like scheduling~\cite{wang2024survey, fuefficient}. More direct applications involve LLMs in the decision-making process for resource allocation. For instance, LLMSched~\cite{zhu2025llmsched} proposes an uncertainty-aware scheduling framework for compound LLM applications themselves, highlighting the complexity LLMs introduce as workloads. Enhanced reasoning in LLMs employs reinforcement learning for intermediate step optimization via internal search or path exploration. Reliability techniques include self-correction~\cite{ferraz2024llm} and contrastive methods, while applications span structured planning model extraction~\cite{tantakoun2025llms} and multiagent coordination.
Unlike previous focus on NLP tasks, our work directly applies LLM reasoning to scheduling decisions through constraint-aware prompting.

\savevspace
\subsection{Reasoning LLMs}
\label{sec:llm}
State-of-the-art LLMs employ ``slow thinking'' mechanism for complex problem-solving:
\begin{itemize}
  \item \textbf{OpenAI's O4-Mini}: RL-trained for multi-step reasoning, excelling in mathematics and coding benchmarks~\cite{openai2025learning}
  \item \textbf{Claude 3.7 Sonnet}: Hybrid architecture offering speed/quality tradeoffs with 200K token context~\cite{wandb2025evaluating}
\end{itemize}

\begin{figure}[htbp]
    \centering
    \includegraphics[width=\linewidth]{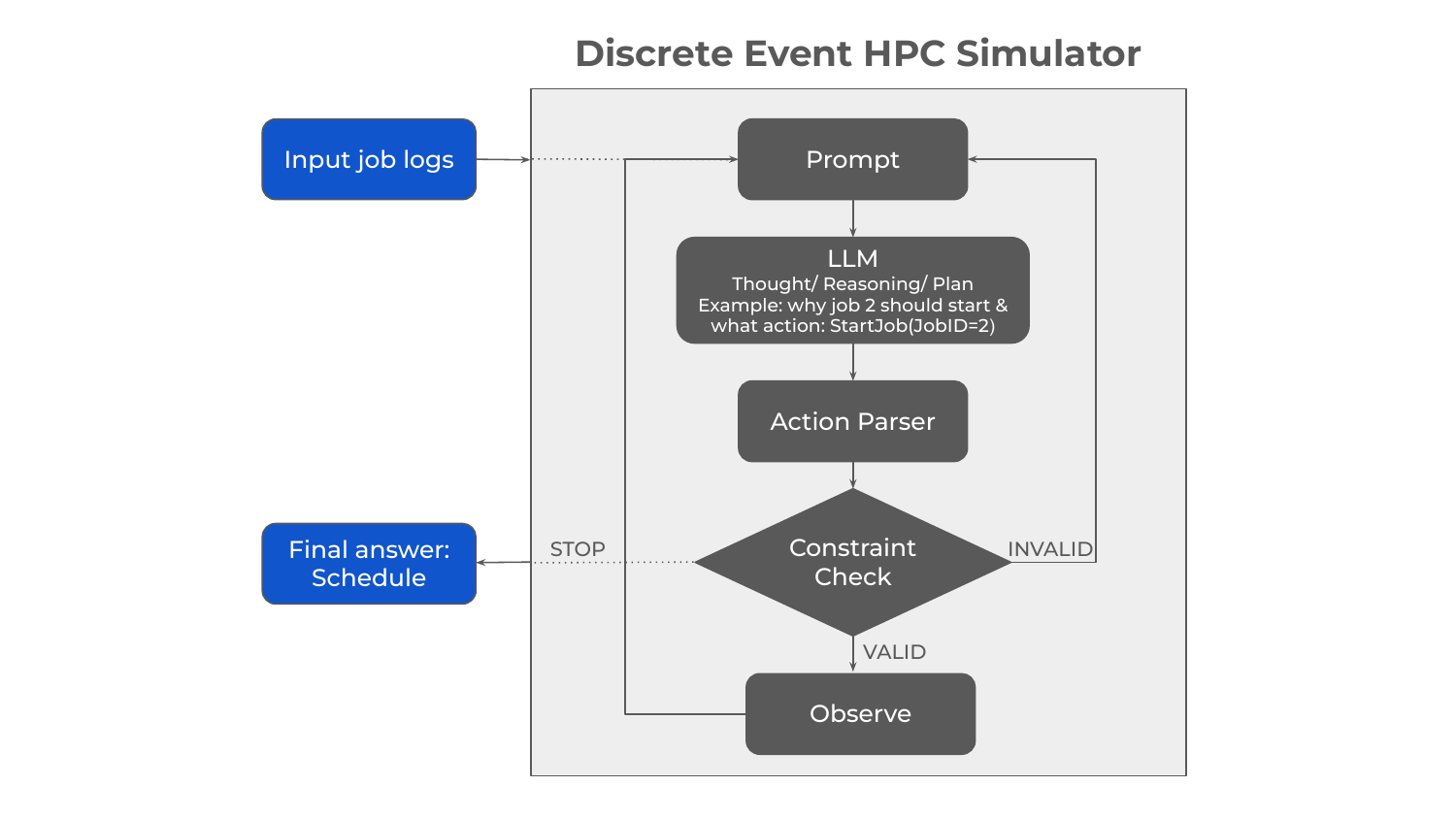}
    \caption{LLM-based HPC scheduling with ReAct-style reasoning.}
    \label{fig:react-hpc}
\end{figure}

While LLMs show success on static benchmarks, HPC scheduling presents dynamic challenges with noisy inputs~\cite{naghshnejad2020hybrid}. Our framework addresses this through: 1) ReAct (reasoning + action) prompting, shown in Figure~\ref{fig:react-hpc}, for constraint-aware decisions, 2) Scratchpad memory maintaining scheduling context, and 3) Natural language feedback for constraint violations. This enables zero-shot adaptation to diverse workloads while ensuring feasibility through modular constraint enforcement.
\savevspace
\section{LLM-based reasoning agent for multiobjective job scheduling}
\label{sec:method}
We present a language-model-driven scheduling agent designed to solve the HPC job scheduling problem through iterative, interpretable reasoning. We refer to our system as an agent because it operates autonomously in a closed-loop interaction with the HPC environment: observing the system state, reasoning over multiobjective goals, and issuing actions to schedule jobs. Unlike traditional schedulers, this agent learns from feedback and evolves its behavior over time, embodying key traits of intelligent agent-based systems. To ensure correctness and realism, the agent interacts with a discrete event simulator, which functions as the underlying HPC environment. The simulator maintains the global simulation clock, tracks system resources, enforces scheduling constraints, and executes valid agent-proposed actions.
This section details the problem formulation, the architecture of the \emph{ReAct-style agent}, and the overall decision-making loop.
\savevspace
\subsection{Problem Formulation}

We define the HPC scheduling problem as a sequential decision-making process. At each timestep $t$, the agent receives a snapshot of the system state $S_t$, a job queue $Q_t = \{j_1, \dots, j_n\}$, where each job $j_i$ has resource demands $r_i = (n_i, m_i)$ for nodes and memory, and a cumulative reasoning history (``scratchpad'').
%
%
The goal is to construct a schedule $\pi = \{x_1, x_2, \dots,  x_n\}$ that assigns start times to jobs while optimizing multiple objectives, including makespan, resource utilization, fairness, machine throughput, subject to the system node and memory capacity, and job feasibility and eligibility constraints.


\savevspace
\subsection{ReAct-style LLM Scheduling Agent}

To apply large language models (LLMs) in a domain like HPC scheduling--characterized by numeric constraints and dynamic job arrival--we use a prompting framework called \emph{ReAct (Reason + Act)}.
ReAct ~\cite{yao2023react} was introduced to enable LLMs to solve complex problems by alternating between:
\begin{itemize}
  \item \textbf{Thought:} a free-form reasoning step where the model explains its logic,
  \item \textbf{Action:} a structured command representing a concrete scheduling decision.
\end{itemize}

This design allows the model to break down complex trade-offs (e.g., fairness vs. throughput) in human-readable terms. In our system, the ReAct agent is prompted with the combination of the current system state (e.g., available nodes, memory), the job queue with metadata, and a running scratchpad that logs all past thoughts, actions, and feedback.
%
%

The model selects from \emph{action spaces}:
\begin{itemize}
  \item \texttt{StartJob(job\_id=X)}: Start job X immediately,
  \item \texttt{BackfillJob(job\_id=Y)}: Opportunistically run a smaller job earlier,
  \item \texttt{Delay}: Wait and defer action until conditions change,
  \item \texttt{Stop}: End the scheduling process.
\end{itemize}

This scratchpad-based prompting acts as a form of memory, enabling continuity across steps without retraining or fine-tuning.
\savevspace
\subsection{Reasoning Loop}

The ReAct prompting framework supports a structured decision loop in Algorithm~\ref{algo:llm_agent}:
\begin{algorithm}
\caption{LLM Scheduling Agent Loop}
\label{algo:llm_agent}
\begin{algorithmic}[1]
  \State \textbf{Input:} Job queue $Q$, initial state $S_0$
  \State Initialize scratchpad $P \gets \emptyset$
  \While{not all jobs are scheduled}
    \State Construct prompt with $S_t$, $Q_t$, and $P$
    \State Query LLM $\rightarrow$ (Thought, Action)
    \If{Action is valid}
      \State Update system state
    \Else
      \State Generate feedback
    \EndIf
    \State Append Thought, Action, and Feedback to $P$
  \EndWhile
\end{algorithmic}
\end{algorithm}

This loop enables correction through natural language feedback and requires no model retraining. 
The LLM adapts in a zero-shot manner using only the structured prompt context.
\savevspace
\subsection{Constraint Enforcement}

Traditional HPC schedulers hard-code feasibility checks. LLMs, by contrast, can hallucinate or misinterpret numeric limits. To ensure validity: the simulator validates each LLM-suggested action; feasible actions are executed; invalid ones are rejected; violations (e.g., memory overflow) are explained in natural language; these explanations are appended to the scratchpad to inform future decisions.
This separation of reasoning from enforcement enables reliable LLM control in HPC environments.

\savevspace
\section{Experiments}
\label{sec:experiments}

\subsection{HPC jobs instance generation}

To emulate diverse and realistic HPC workloads, we define seven benchmark scenarios, each reflecting a distinct operational pattern observed in real-world job traces. 
The workloads are synthetically generated using a scenario-driven simulator that produces job logs with varying runtime distributions, resource demands, and user metadata. 
This simulation approach enables precise control over workload characteristics while capturing the variability seen in production HPC environments.

Jobs arrive dynamically over time, following workload-specific interarrival patterns. To mimic realistic submission behaviors in HPC environments, we simulate job arrival times using Poisson processes, commonly used to model inter-arrival patterns in real-world queuing systems. For each workload scenario, we define a scenario-specific arrival rate 
$\lambda$, which governs the average time between job submissions. This dynamic queueing captures timing variability and contention, enabling a more realistic evaluation of scheduling strategies. The simulated HPC environment models a cluster partition with 256 compute nodes and 2048 GB of total memory capacity, representing typical shared-resource configurations found in scientific computing centers. Each submitted job specifies node counts, memory requirements, and estimated walltime, executing non-preemptively once resources become available and scheduling decisions are made. 
The simulator operates as a discrete event system, advancing simulation time only at key events such as job arrivals and job completions. At each step, the simulator injects any newly arrived jobs into the waiting queue, updates the status of running jobs (releasing resources for those that have finished), and then determines the next scheduling action. If there are jobs ready to be scheduled, the agent queries the LLM for a decision; otherwise, it advances time to the next event (either the next job arrival or the next job completion). This event-driven approach ensures that the simulation efficiently models the dynamic scheduling environment without unnecessary time increments, accurately reflecting real-world system behavior where state changes occur only at discrete points in time.

We evaluate scheduling performance across seven dynamically generated, each reflecting distinct HPC usage patterns and arrival behaviors: \textbf{Homogeneous Short} (uniform 30-120s jobs with 2 nodes, 4GB) simulates lightweight CI/test workloads; \textbf{Heterogeneous Mix} features varied runtime (Gamma distribution, shape=1.5, scale=300) and resource requirements reflecting realistic production environments; \textbf{Long-Job Dominant} combines 20\% extremely long jobs (50,000s, 128 nodes) with short jobs (500s, 2 nodes) to test convoy effect handling; \textbf{High Parallelism} uses large parallel jobs (64-256 nodes, Gamma walltime distribution) simulating tightly-coupled simulations; \textbf{Resource Sparse} employs lightweight jobs (1 node, $<$8GB, 30-300s) testing sparse workload efficiency; \textbf{Bursty + Idle} alternates between short and long-running jobs with modest resource demands to evaluate responsiveness under uneven job durations; and \textbf{Adversarial} stress-tests schedulers with one large blocking job (128 nodes, 100,000s) followed by many small jobs (1 node, 60s) to expose convoy effects.

Each scenario was instantiated with [10, 20, 40, 60, 80, 100] jobs. We collect metrics including makespan, average wait time, turnaround time, node/memory utilization, throughput, and fairness to compare scheduler performance under these evolving conditions.
\savevspace
\subsection{Objectives}

We evaluate scheduling performance using seven standard objectives that capture both system-level efficiency and user-perceived responsiveness. \textbf{Makespan}~\cite{englert2008power} measures the total elapsed time from earliest job submission to completion of the last job ($\max_{j \in J} (x_j + d_j)$), reflecting overall schedule efficiency. \textbf{Average wait time}~\cite{smith1999using} captures user-perceived latency as the time jobs spend queued before execution ($w_j = x_j$ since all $s_j = 0$), while \textbf{average turnaround time}~\cite{smith1999using} measures the total time from submission to completion ($x_j + d_j$). \textbf{Throughput}~\cite{mekkittikul1998practical} indicates queue processing speed as jobs completed per unit time ($n/(\text{Makespan} - \min_{j}(x_j))$). \textbf{Resource utilization}~\cite{lee2014improving} is measured separately for nodes ($\frac{1}{C\cdot \text{Makespan}}\sum_{j \in J}n_j\cdot d_j$) and memory ($\frac{1}{M \cdot \text{Makespan}}\sum_{j \in J}m_j\cdot d_j$), where $C$ and $M$ represent total system capacity and $n_j$, $m_j$ are per-job resource requirements. \textbf{Fairness}~\cite{ajtai1998fairness} is evaluated using Jain's index~\cite{sediq2013optimal} from both per-job ($\frac{(\sum_{j \in J} w_j)^2}{n \cdot \sum_{j \in J} w_j^2}$) and per-user ($\frac{(\sum{i \in U} u_i)^2}{|U| \cdot \sum_{i \in U} u_i^2}$) perspectives, where $u_i$ represents average wait time per user, with values ranging from 0 (worst) to 1 (perfect fairness).

\savevspace
\subsection{Experimental  setup}
We evaluated our LLM-based scheduler using two models: O4-Mini (OpenAI, hosted on Microsoft Azure) and Claude 3.7 (Anthropic, accessed via Google Cloud Vertex AI). LLMs were accessed via APIs in a single-turn ReAct-style prompting loop at each discrete step. For O4-Mini, we used the ``reasoning effort: high'' setting with a maximum token context of 100,000. The temperature was fixed internally and could not be controlled. For Claude 3.7, we used a maximum of 5,000 tokens with temperature = 0 to ensure deterministic outputs.

Given $n$ jobs $J = \{1, 2, \ldots, n\}$ submitted simultaneously at time $s_j = 0$ (i.e., all jobs arrive at system initialization), each job $j \in J$ is characterized by three key parameters $d_j, n_j, m_j$ as job duration, nodes requested, and memory (in GB) requested for job $j$, respectively.
The LLM-based scheduler must determine start times $\{x_j\}_{j=1}^n$ for all jobs while optimizing system objectives (e.g., makespan minimization, resource utilization maximization) under the constraints:
$\sum_{j \in A(t)} n_j \leq N_{\text{total}}$ (Node capacity), $\forall t$
$\sum_{j \in A(t)} m_j \leq M_{\text{total}}$ (Memory capacity), $\forall t$
where $A(t) = \{j \in J \mid x_j \leq t < x_j + d_j\}$ denotes the set of active jobs at time $t$, and $N_{\text{total}}$, $M_{\text{total}}$ represent the cluster's total node and memory resources respectively.

We compare our LLM-based scheduler to the following methods: 
First Come First Serve (FCFS) is the simplest scheduling algorithm that executes jobs strictly in their arrival order, subject to resource constraints.
Shortest Job First (SJF) prioritizes jobs with the shortest estimated runtime, typically reducing average turnaround time but potentially starving longer jobs, compromising fairness. 
Google OR-Tools~\cite{ortools} provides an optimization-based scheduling solution, which we use as a strong baseline; it computes globally optimal or near-optimal schedules for small-to-medium workloads, offering a performance upper bound for comparison. 
We do not use reinforcement learning (RL)\cite{wang2021rlschert} or graph neural networks (GNNs)~\cite{adimora2024gnn} in this work, as these methods typically optimize for a single scalar objective (e.g., reward or cost) and require extensive environment tuning and retraining for new objective trade-offs. In contrast, our LLM-based ReAct agent supports multiobjective reasoning in a flexible, interpretable way (without retraining) by conditioning on natural language prompts that describe scheduling goals for jobs.

Our LLM scheduler operates at the job selection and allocation level, using a first-fit strategy on a cluster (256 CPUs, 2048 GB memory). A first-fit strategy allocates each selected job to the first available set of resources that meet its requirements. While we abstract system-level details like topology and storage, these are typically handled by the underlying resource management stack. The focus is on evaluating high-level scheduling decisions, with topology-aware placement left for future work.
All workload data, code, prompts, and experiment configurations are publicly available for reproducibility: \url{https://github.com/swarm-workflows/ReasonedScheduler/tree/master/llm_scheduler}.

\tcbset{
  reasoningbox/.style={
    breakable,
    enhanced,
    colback=gray!5,
    colframe=gray!50!black,
    sharp corners,
    boxrule=0.4pt,
    fonttitle=\bfseries,
    title=LLM Reasoning Trace,
    coltitle=white,
    left=6pt,
    right=6pt,
    top=4pt,
    bottom=4pt,
    boxsep=5pt,
    listing only,
    listing options={basicstyle=\ttfamily\footnotesize,breaklines=true}
  }
}
\savevspace
\subsection{LLM prompt and reasoning traces}
\begin{figure*}
    \includegraphics[width=\textwidth]{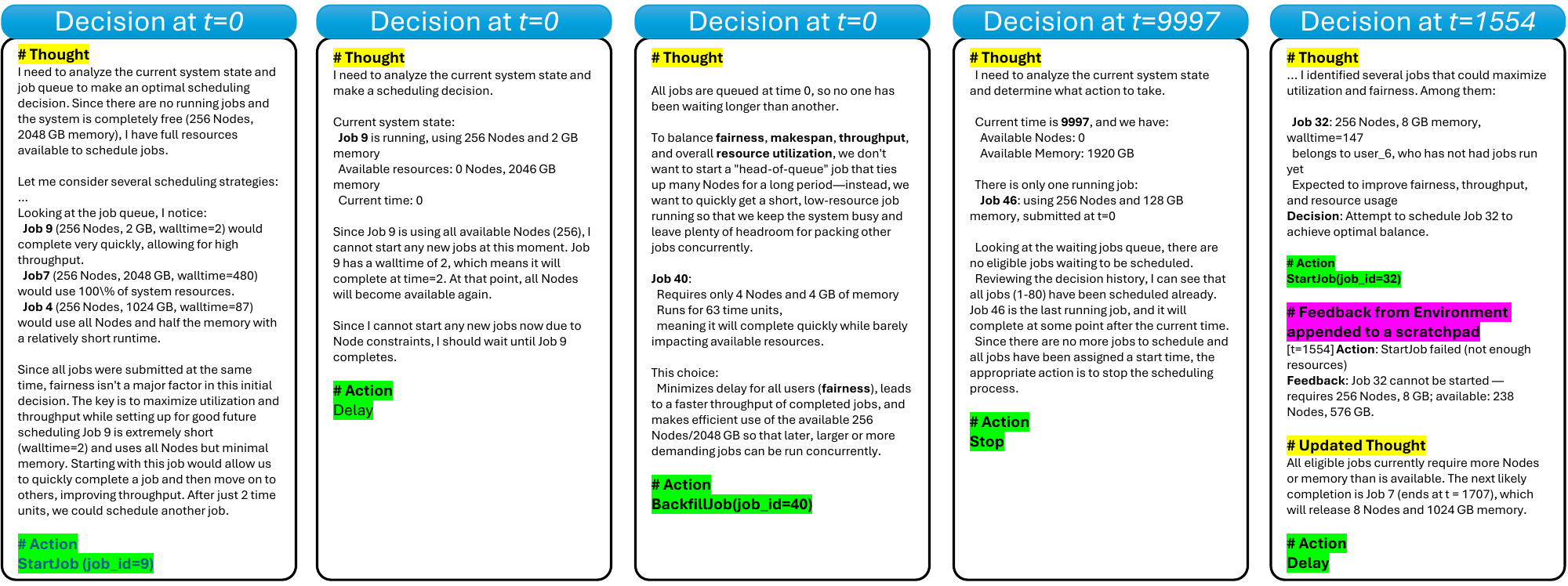}
    \savevspace
\caption{Representative LLM reasoning traces demonstrating interpretable scheduling decisions. The traces show multiobjective trade-off analysis (selecting Job 9 for throughput optimization), resource constraint recognition (delaying when nodes unavailable), intelligent backfilling strategies (choosing Job 40 for efficient resource utilization), and adaptive recovery from constraint violations through natural language feedback. Each decision includes explicit reasoning about fairness, utilization, and system state, providing complete transparency into the scheduling logic. Decision at t = X denotes the simulation time at which the LLM is queried to make a scheduling decision based on the current system state.}

    \label{fig:reasoning_traces}
\end{figure*}

A key feature of our LLM-based scheduler is that all scheduling decisions are made through text-based reasoning, providing full visibility into the model’s thought process. This enhances interpretability and enables qualitative analysis of the LLM’s emergent heuristics and decision strategies.
Below, we highlight the prompt that we have constructed  and show the excerpts from representative decision traces in  Figure~\ref{fig:reasoning_traces} observed during experiments.

\tcbset{
  promptbox/.style={
    enhanced,
    colback=blue!5!white,           
    colframe=blue!60!black,         
    colbacktitle=blue!20!white,     
    coltitle=black,                 
    fonttitle=\bfseries\large,
    fontupper=\footnotesize\ttfamily,
    title={\textbf{Prompt Example}}, 
    boxrule=0.8pt,
    sharp corners,
    breakable,
    top=5pt,
    bottom=5pt,
    left=5pt,
    right=5pt,
    boxsep=5pt
  }
}

\begin{tcolorbox}[promptbox]
You are an expert HPC resource manager, and your task is to schedule jobs in a high-performance computing (HPC) environment. Use the current system state, job queue, scratchpad (decision history), and fairness indicators to make well-balanced decisions.

\textbf{\textcolor{blue!80!black}{System capacity}}: 256 nodes, 2048 GB memory \\
\textbf{\textcolor{blue!80!black}{Current time}}: 0 \\
\textbf{Available Nodes}: \texttt{256} \\
\textbf{Available Memory}: \texttt{2048 GB}

\textbf{Running Jobs:} \\
\texttt{None}

\textbf{Completed Jobs:} \\
\texttt{None}

\textbf{Waiting Jobs (eligible to schedule):} \\
\texttt{None}

\textbf{\# Scratchpad (Decision History)} \\
\texttt{(nothing yet)}

\textbf{\textcolor{blue!80!black}{Your scheduling objectives are:}} \\
You must balance \textbf{all} of the following:
\begin{itemize}
  \item \textbf{Fairness}: Minimize variance in user wait times. Avoid starving any user.
  \item \textbf{Makespan}: Minimize total time to finish all jobs.
  \item \textbf{Utilization}: Maximize Node \& memory usage over time (avoid idle resources).
  \item \textbf{Throughput}: Maximize the number of jobs completed per unit time.
  \item \textbf{Feasibility}: Do not exceed 256 Nodes or 2048 GB memory at any time.
\end{itemize}

\textbf{\textcolor{gray!70!black}{Trade-offs are allowed.}} Do not over-optimize one metric at the expense of others. \\
For example:
\begin{itemize}
  \item Prioritizing a long-waiting job improves fairness, but may slightly hurt makespan.
  \item Choosing short jobs improves throughput, but may increase wait time for large jobs.
\end{itemize}

\textbf{\textcolor{blue!80!black}{Decide:}}
\begin{enumerate}
  \item Which job should be started now (if any)?
  \item Justify your decision in thought.
  \item Return only one of:
    \begin{itemize}
      \item \texttt{StartJob(job\_id=X)}
      \item \texttt{BackfillJob(job\_id=Y)}
      \item \texttt{Delay}
      \item \texttt{Stop} (when all jobs have been scheduled)
    \end{itemize}
\end{enumerate}

\textbf{\textcolor{blue!80!black}{Output format:}} \\
\texttt{Thought: <your reasoning>} \\
\texttt{Action: <your action>}
\end{tcolorbox}

\savevspace
\subsection{Comparison across diverse workload scenarios}
\begin{figure*}[t]
    \centering
    \includegraphics[width=0.9\textwidth]{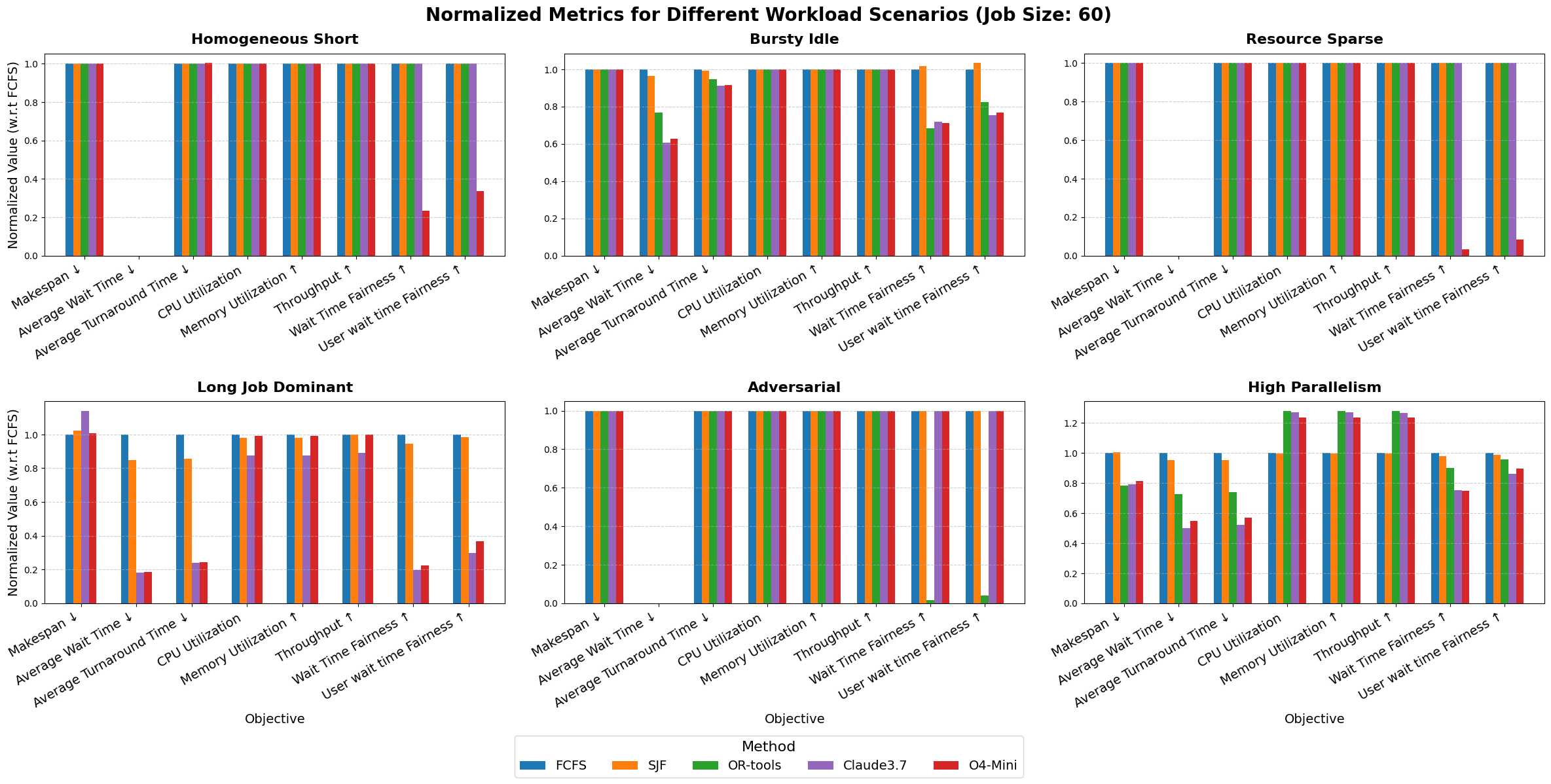}
    \savevspace
\caption{Normalized performance metrics across six workload scenarios with 60 jobs each. All metrics are normalized relative to FCFS performance. Lower values indicate better performance for negative metrics (makespan, wait \& turnaround times), while higher values indicate better performance for positive metrics (utilization, throughput, and fairness). The comparison includes traditional heuristics (FCFS, SJF), optimization-based scheduling (OR-Tools), and LLM-based approaches. Claude 3.7 demonstrated superior adaptability and multiobjective balance in Bursty Idle and Long Job Dominant. In contrast, OR-Tools focuses on utilization but degrades in fairness, while heuristics show limited responsiveness to workload variation.}
    \label{fig:react-exp2}
\end{figure*}
%

We evaluated our LLM-based scheduling approach against traditional baselines across seven workload scenarios with 60 jobs each, with performance metrics normalized relative to FCFS (baseline=1.0). The results, shown in the normalized metrics chart (Figure~\ref{fig:react-exp2}), reveal distinct performance patterns that highlight the strengths and limitations of different scheduling approaches under varying workload characteristics.
Note that the figure did not include Heterogeneous Mix because it will be covered in the section \ref{sec:scalability}.Also, the average wait time for all methods, including FCFS, was 0.0. Since the normalization is performed relative to FCFS, the resulting value becomes undefined (0/0) and is therefore omitted from the comparison.

\textbf{Bursty Idle} highlights the limits of traditional and optimization-based schedulers. OR-Tools exhibits a degradation in wait time fairness, while both LLM-based schedulers maintain balanced performance across all objectives, including throughput, fairness, and utilization, demonstrating adaptability under workload volatility.
\textbf{Long Job Dominant} shows pronounced differences. FCFS, SJF suffer significantly from the convoy effect, with higher wait time, turnaround time. LLMs showed dramatically reduced average wait and turnaround time. O4-mini performs very well with balanced fairness and efficiency. Claude 3.7 performed slightly worse on fairness. 
\textbf{In High Parallelism}, OR-Tools and LLMs achieve the highest resource utilization and throughput (>1.2x), benefiting from reasoning-based job packing compared with the heuristic methods. O4-Mini performs better than Claude 3.7 in fairness. FCFS, SJF trails behind, indicating limited capacity to handle fragmentation from large job sizes.
\textbf{Adversarial} conditions lead to flattened differences: all methods perform nearly identically across metrics. LLMs like Claude 3.7 and O4-Mini perform no worse than heuristics ones; a good sign of stability. OR-tools achieved optimal or near-optimal results in throughput, turnaround time, and system utilization across scenarios. However, this gain came at the cost of fairness: both wait time fairness and user-level fairness were lower compared to simpler heuristics like SJF and FCFS, indicating potential bias in scheduling priority or resource allocation. 
\textbf{Resource Sparse} again yields limited differentiation; all schedulers perform near the baseline, reflecting minimal contention. LLMs don’t gain much advantage here, there’s little room for intelligent scheduling to make an impact. O4-Mini showed notably poor fairness. Its learned policy likely optimize for system-wide efficiency, prioritizing “easy wins” (e.g., smaller jobs). This causes starvation or long delays for certain jobs or users when resources are scarce.
\textbf{Homogeneous Short} scenarios also result in uniform performance. All methods deliver similar makespan, throughput, and fairness, as the low variability in job characteristics reduces the need for strategic scheduling. O4-Mini shows reduced fairness, likely due to job similarity making simple methods sufficient.

\savevspace
\subsection{Scalability analysis}
\label{sec:scalability}

\begin{figure*}[ht]
    \centering
    \includegraphics[width=0.9\textwidth]{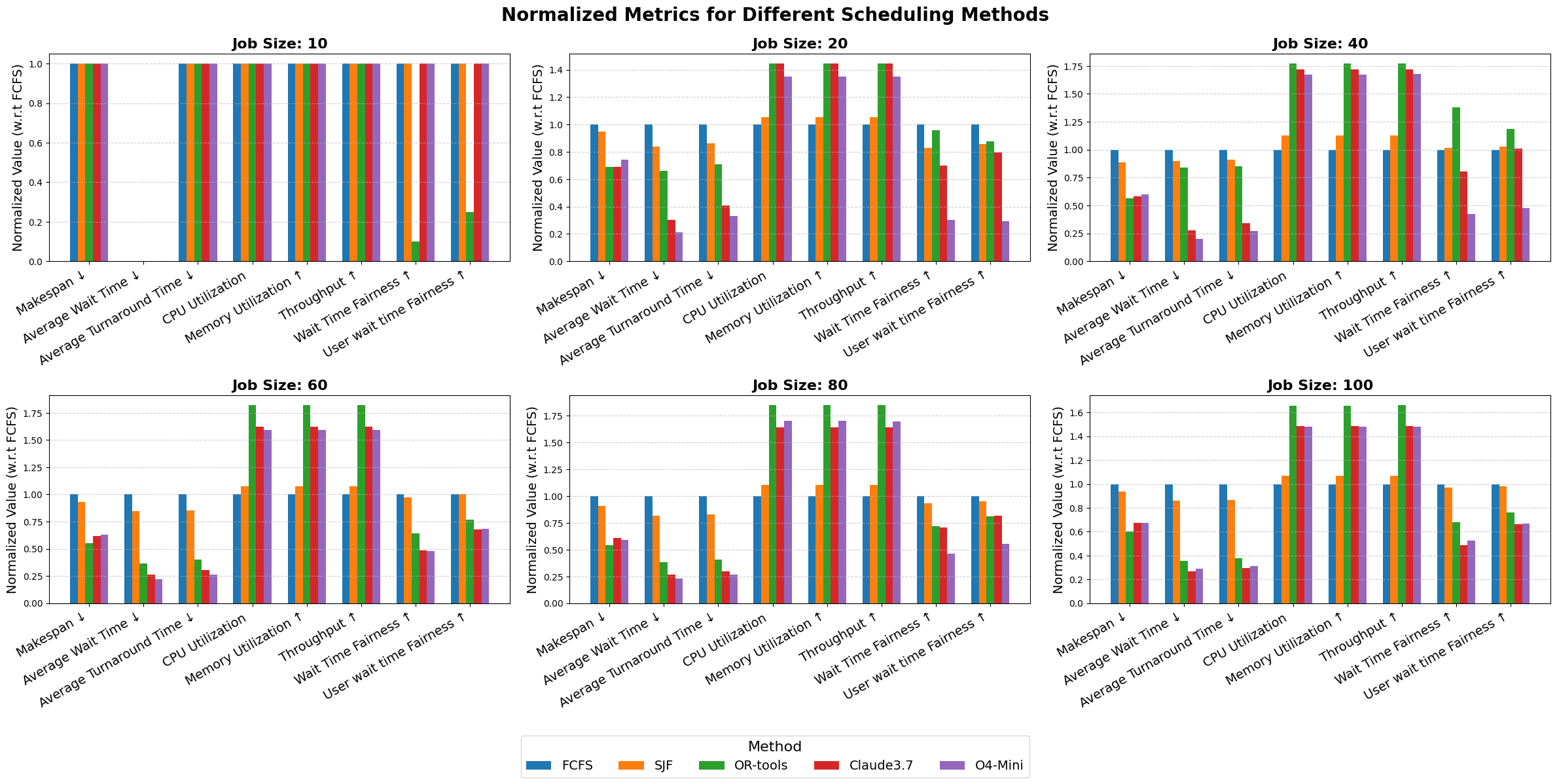}
    \savevspace
\caption{Scalability analysis showing normalized performance metrics across increasing job queue sizes (10 to 100 jobs) for the Heterogeneous Mix workload scenario. All metrics are normalized relative to FCFS performance (baseline = 1.0). The results demonstrate how scheduling performance differentiates as problem complexity increases. At small scales (10 jobs), all methods perform similarly, but significant performance gaps emerge with complexity. OR-Tools achieves exceptional resource utilization (up to 1.75× FCFS) at higher scales but shows more variable performance across other metrics. LLM-based schedulers maintain consistently balanced performance across all metrics and scales, demonstrating robust scalability characteristics. Traditional heuristics (FCFS, SJF) show limited improvement potential as queue size increases, highlighting the importance of advanced scheduling approaches for large-scale HPC workloads.}
\label{fig:scalability-analysis}
\end{figure*}

To evaluate how scheduling performance scales with problem complexity, we analyzed the Heterogeneous Mix workload scenario across increasing job queue sizes from 10 to 100 jobs. The results shown in Fig.~\ref{fig:scalability-analysis} demonstrate a clear relationship between problem scale and the effectiveness of different scheduling approaches, with performance differentiation becoming increasingly pronounced as queue size grows.
As shown in Figure 4, performance differentiation becomes more pronounced as job size increases:
Small queues (10–20 jobs): All methods perform comparably across most objectives, indicating low contention and reduced room for optimization. OR-Tools slightly improves utilization early on, while LLM-based agents start to show fairness benefits.
Mid-scale queues (40–60 jobs): Claude 3.7 begins to outperform others across multiple metrics, particularly in wait time, turnaround time, and fairness at the same time. This suggests emerging benefits from reasoning-based adaptation as scheduling complexity grows. OR-Tools continues to optimize resource utilization but degrades in wait time and turnaround time.
Large queues (80–100 jobs): Performance differences widen further. Claude 3.7 and O4-Mini maintain multiobjective balance, with strong throughput (1.4x) and resource utilization (>1.4x vs FCFS) while still achieving fairness. In contrast, OR-Tools maximizes CPU and memory utilization (up to 1.8×) but shows higher wait time \& turnaround time, reflecting a trade-off due to its optimization focus. FCFS and SJF remain largely static across scales, showing limited adaptability.
This analysis highlights the graceful scalability of LLM-based schedulers, which maintain balanced decision-making across job sizes, contrasting with the rigidity of heuristics and the singular optimization focus of OR-Tools. Reasoning-based agents dynamically adapt to queue complexity and maintain user-centric objectives under load.

\savevspace
\subsection{Computational Overhead analysis}

\begin{figure*}[htbp]
\centering
\includegraphics[width=0.9\textwidth]{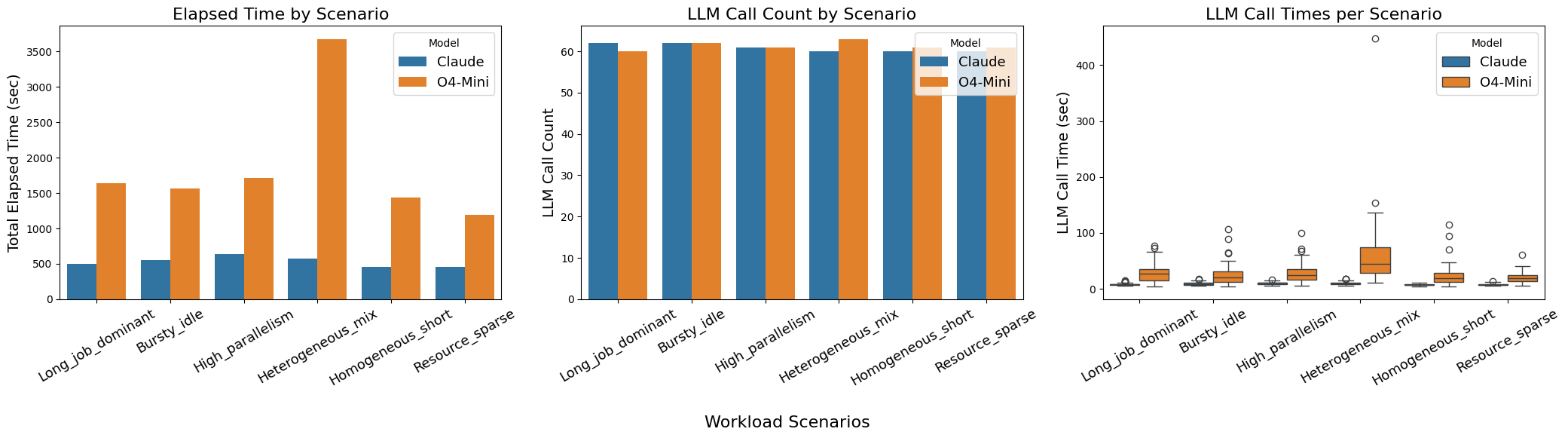}
\caption{Computational overhead comparison across six workload scenarios. Metrics include total elapsed time (left), number of LLM calls (center), and per call latency distribution (right) for Claude 3.7 and O4-Mini. Only successful scheduling actions (start\_job, backfill\_job) are considered. Claude 3.7 demonstrates significantly lower and more stable inference overhead, especially in complex workloads like Heterogeneous Mix, where O4-Mini exhibits high latency variance.}
\label{fig:llm-efficiency}
\end{figure*}

\begin{figure*}[htbp]
\centering
\includegraphics[width=0.9\textwidth]{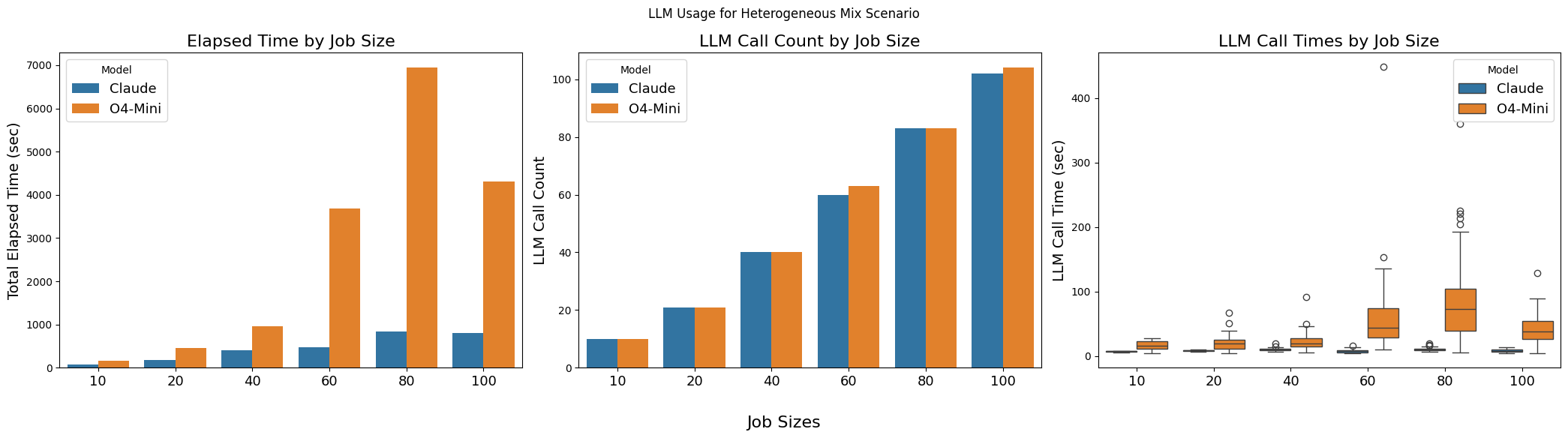}
\caption{LLM computational Overhead scaling with job queue size for Heterogeneous Mix workloads. Scheduling time grows super-linearly, reaching 4k s (O4-Mini) and 700 s (Claude 3.7) for 100 jobs, while API calls scale linearly. O4-Mini exhibits a pronounced spike in total runtime at job size 80, exceeding 6.9k s, despite having fewer jobs than the 100-job case. It could be the case that the workload generation process or attributable to transient network/API latency during execution, rather than purely scheduling-reasoning complexity.}
\label{fig:llm-scalability}
\end{figure*}

A critical consideration for deploying LLM-based scheduling in production HPC environments is understanding the computational overhead and how it scales with both workload complexity and problem size. We analyze the overhead characteristics across two dimensions: workload-dependent complexity at fixed scale (60 jobs) and scalability with increasing job queue sizes for the most representative scenario.
\subsubsection{Workload-dependent overhead characteristics.}
To ensure fair and meaningful overhead analysis, we restrict our measurements to LLM calls that resulted in feasible and accepted scheduling actions; namely, \textit{start\_job} and \textit{backfill\_job}. Calls leading to delay events are excluded, as they typically suggested either when the system is fully saturated (i.e., all resources are occupied), or when pending jobs demand more resources than currently available, making any scheduling action infeasible. Including such cases would conflate LLM reasoning latency with external system constraints. 
Thus, our analysis isolates the LLM’s decision overhead under conditions where successful job placement is possible.
For each workload model pair, we collected the total elapsed scheduling time, number of LLM calls, and the distribution of individual LLM call latencies. The plot below in Figure \ref{fig:llm-efficiency} compares these overhead metrics between Claude 3.7 and O4-Mini.
Elapsed Time (Left Plot): Claude 3.7 consistently exhibits lower total elapsed scheduling time across all workload types compared to O4-Mini, with up to 7× faster execution for Heterogeneous Mix. This highlights Claude 3.7's faster end-to-end decision efficiency. LLM Call Count (Middle Plot): All models exhibit ~60 calls (equal to job count), with slight variation due to backfilling. This confirms comparable action trace lengths, indicating that runtime differences stem from per-call latency, not the number of decisions. And LLM Call Time Distribution (Right Plot): Claude 3.7’s per-call latencies are tightly clustered below 10 seconds, showing low variance and stable reasoning latency. In contrast, O4-Mini exhibits high variance, with several outliers exceeding 100s, particularly in Heterogeneous Mix. This behavior indicates struggles in complex state-action mappings requiring longer reasoning chains.
The elevated overhead observed for O4-Mini in the Heterogeneous Mix scenario can be attributed to the increased complexity of decision-making under resource heterogeneity and scheduling contention. In this setting, jobs vary widely in size (nodes, memory) and duration, creating a larger and more constrained scheduling search space. The observed latency patterns are consistent with increased reasoning difficulty driven by workload diversity.

\subsubsection{Scalability with job queue size}
The scalability analysis for Heterogeneous Mix workloads reveals how computational overhead grows with problem size, as depicted in Figure~\ref{fig:llm-scalability}. This analysis is crucial for understanding deployment feasibility in large-scale HPC environments.

Total Elapsed Time (Left): Both models show a monotonic increase in total elapsed time as job size increases, but O4-Mini’s growth is superlinear from 40 jobs onward. This suggests compounding latency effects, potentially due to increased reasoning complexity when the waiting queue contains more heterogeneous jobs. Claude 3.7 scales more gracefully, maintaining a near-linear growth profile.
LLM Call Count (Middle): Call counts scale proportionally with job size for both models, as more scheduling decisions are needed. The similarity in call counts indicates that elapsed time differences stem from per call latency, not from a larger number of decisions. LLM Call Times Distribution (Right): Claude 3.7 maintains a tight latency distribution across job sizes, with minimal outliers. O4-Mini exhibits significant spread and heavy-tailed latency distributions, especially at 60–80 jobs, where outliers exceed 200 s. This variance indicates occasional very slow decision episodes, likely caused by increased search space and reasoning over conflicting scheduling objectives.




\subsubsection{Deployment implications}

The overhead analysis reveals several critical considerations for deploying LLM-based scheduling in production HPC environments. The wall-clock times required (up to an hour for 100 jobs) indicate that, at the moment, LLM-based scheduling is not suitable for real-time job submission scenarios. However, it may be practical for batch scheduling systems, periodic resource optimization, or strategic scheduling decisions where quality is prioritized over immediate response.



For HPC installations regularly processing hundreds or thousands of concurrent jobs, the computational overhead may become prohibitive. The analysis suggests practical deployment limits around 100-200 jobs for periodic scheduling optimization, beyond which the overhead may outweigh the scheduling benefits.

The results suggest that on-premise fast reasoning models are critical to overcome the computational overhead barriers identified in this analysis. As reasoning model efficiency continues to improve and deployment costs decrease, LLM-based scheduling may become viable for increasingly latency sensitive and large-scale HPC applications.

%
%





\begin{figure*}[htbp]
\centering
\includegraphics[width=0.9\textwidth]{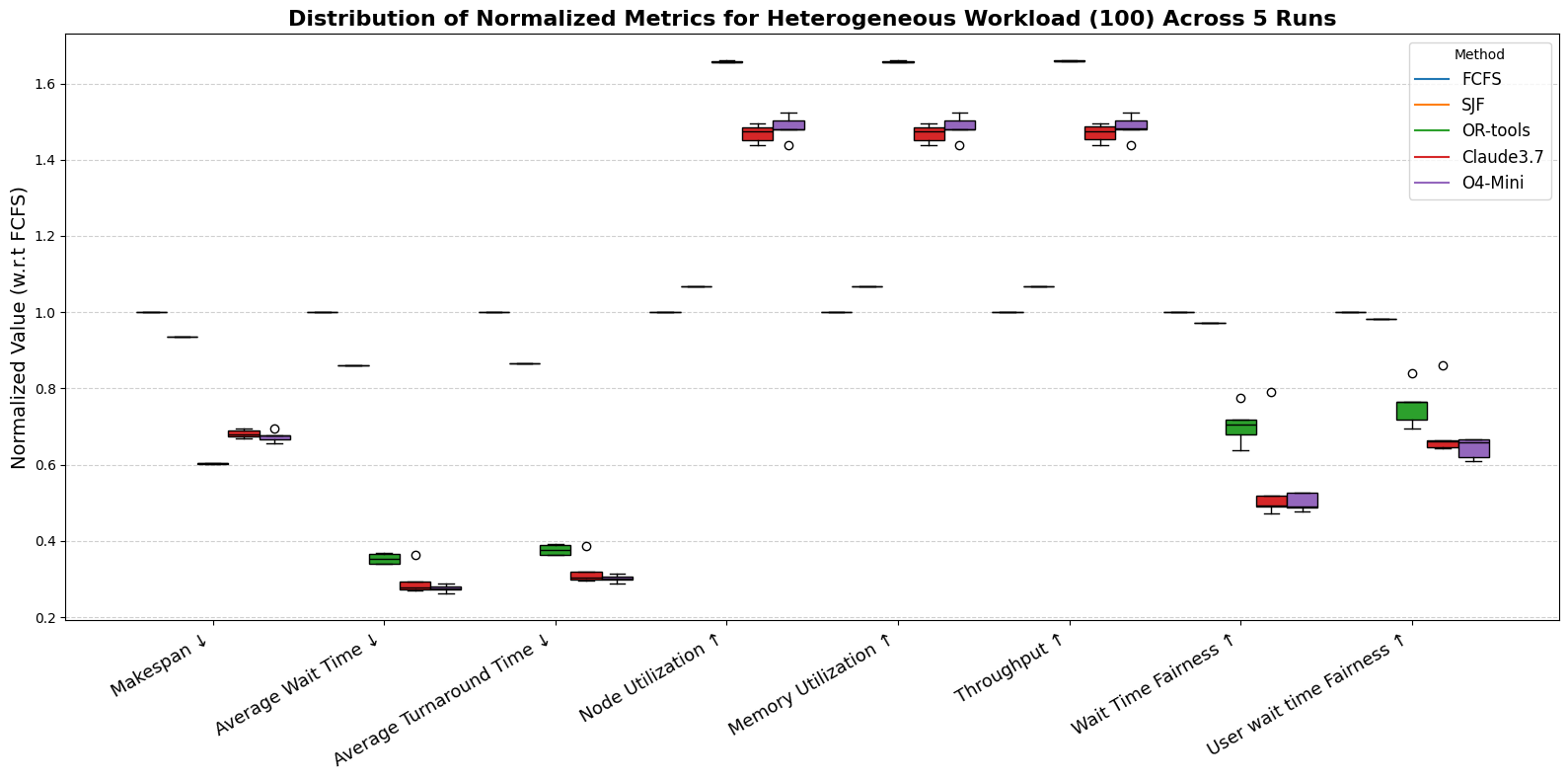}
\caption{Distribution of normalized perofrmance metrics for the Heterogeneous Mix workload with 100 jobs over 5 independent runs, normalized to FCFS (baseline=1.0). LLM-based schedulers (Claude 3.7, O4-Mini) show low variance and consistent improvements across multiple objectives, while OR-Tools excels in utilization but exhibits fairness trade-offs. FCFS and SJF remain deterministic with static performance. Results demonstrated the robustness of LLM scheduling under stochastic reasoning.}
\label{fig:stat}
\end{figure*}

\section{Statistical Robustness and Repeatability}
Given the inherent stochasticity of LLMs, we evaluate the statistical robustness of our results through multiple independent runs of each scheduling method. Specifically, we conduct 5 repetitions of the full scheduling pipeline for the Heterogeneous Mix workload with 100 dynamically arriving jobs (We selected the Heterogeneous Mix scenario with 100 jobs to evaluate scheduler robustness under realistic, high-load conditions that involve diverse job durations and resource demands—emulating the variability and contention typical in production HPC workloads.), and repeat metric distributions normalized against the FCFS baseline. Figure \ref{fig:stat} presents box plots for eight performance metrics across five methods. Each box represents the distribution of outcomes across five trials per method, providing insight into the stability and variance of each scheduler under repeated execution.
Key observations include: LLM-based schedulers show tight variance bounds for makespan, wait time, turnaround time, and throughput, with consistent improvements over heuristics. Claude 3.7 exhibits strong user-level fairness and low wait times with minimal variance, indicating reliable performance across runs despite non-determinism. OR-Tools achieves the highest utilization but shows greater variance in fairness, revealing a trade-off between resource efficiency and user experience consistency. FCFS and SJF remain deterministic and thus flat in their metric distribution, yet underperform in multiobjective balance. No significant outliers were observed for LLM methods on negative metrics, suggesting robustness in the ReAct-style reasoning framework. 
These results demonstrated that, even without fine-tuning or deterministic control, LLM-based agents produce repeatable and statistically stable scheduling behavior under complex, dynamic workloads with the lowest temperature parameter. This supports the viability of using zero-shot LLMs for multiobjective HPC scheduling, especially when transparency and fairness are critical.

\section{Evaluation on real workload traces}
\begin{figure*}[htbp]
\centering
\includegraphics[width=0.9\textwidth]{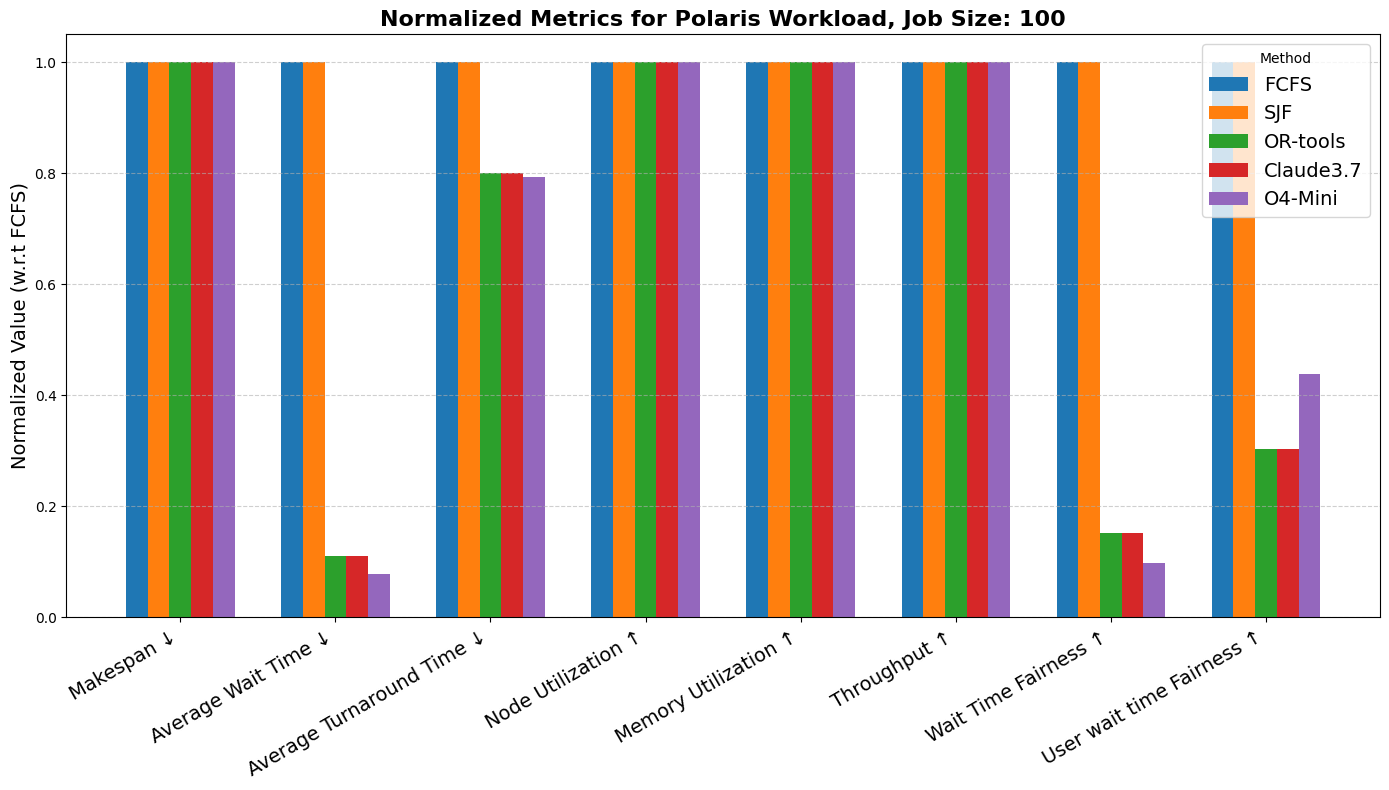}
\caption{Normalized performance metrics for 100  Polaris job logs. LLM-based schedulers reduce latency and improve fairness while maintaining system efficiency.}
\label{fig:real_workload}
\end{figure*}

To assess the applicability of our LLM-based scheduler on real-world workloads, we evaluated it using job traces from the Polaris supercomputer at Argonne National Laboratory. Polaris consists of 560 compute nodes, each with 512 GB of memory. Polaris data was generated from resources of the Argonne Leadership Computing Facility, which is a DOE Office of Science User Facility supported under Contract DE-AC02-06CH11357. We preprocessed 100 jobs submitted in November 2024 from the public job history logs, selecting a contiguous segment of completed jobs for evaluation. Data Preprocessing: The raw trace was filtered to remove failed jobs (i.e., $EXIT\_{STATUS} = -1$) and sorted by submission time. Key fields such as job name, user, group, submission, start, and end timestamps, requested nodes, walltime, and queued wait time were retained. Timestamps were normalized relative to the earliest submission to produce relative submission, start, and end times in minutes. Job identifiers, user, and group labels were factorized to anonymized IDs (e.g., User\_1, Group\_1). We used node count as is and derived total memory (using 512 GB per node) per job, yielding a dataset compatible with our LLM agent.

Assumptions and Limitations: Since the trace was sampled mid-stream, the exact system state at the beginning of the sequence is unknown. For simplicity, we assumed the cluster was idle at time zero, an optimistic assumption that deviates from the real system conditions, where nodes may have been partially occupied. Consequently, this setup is not intended as a fair comparison against the actual Polaris scheduler, which operates under complex queue policies and system-wide constraints. Rather, this experiment demonstrates that the LLM scheduler generalizes to real traces and can produce plausible schedules when provided with normalized historical data. 

This evaluation underscores that, even under simplified assumptions, LLM-based scheduling can be grounded in real-world HPC logs, highlighting the generality of the reasoning framework.
Figure \ref{fig:real_workload} presents normalized performance metrics for five schedulers evaluated on 100 real jobs from the Polaris workload trace. All values are normalized for FCFS. The LLM-based schedulers achieve substantial improvements in average wait time and turnaround time, comparable to the idealized SJF policy. Importantly, both LLM agents maintain resource utilization and throughput on par with all baselines, confirming that system efficiency is preserved. 
In terms of wait time and turnaround time, the LLM schedulers show above and at par performance as OR-tools. Claude 3.7, Or-Tools achieves the highest wait time fairness, and O4-Mini demonstrates superior user-level fairness. These results highlight that, even under real-world job traces and an assumed idle system state, LLM agents can generalize effectively and deliver low latency and balanced fairness without explicit optimization objectives.

\savevspace
\section{Conclusion}

We have conducted the first comprehensive evaluation of reasoning-capable Large Language Models for multiobjective HPC job scheduling, demonstrating their potential and limitations through systematic analysis across diverse workload scenarios. Our ReAct-style scheduling agent enables interpretable, zero-shot scheduling that balances makespan, utilization, throughput, and fairness without requiring domain-specific training. The separation of reasoning from constraint enforcement ensures system reliability while providing complete decision transparency essential for HPC operations.

Through evaluation across seven scenarios and varying job sizes, we demonstrate that LLM-based scheduling performs well in complex, heterogeneous environments where traditional heuristics struggle with multiobjective trade-offs. 
However, our computational overhead analysis reveals significant time requirements, with scheduling times ranging from 1-2 hours for 100 jobs using current cloud-based API services. Our evaluation establishes that while current cloud-based LLM services impose significant latency constraints limiting real-time deployment, the consistent multiobjective balancing and superior adaptivity demonstrated by reasoning-based approaches suggest potential for HPC scheduling applications with appropriate deployment strategies.

Our future research directions include extending our approach to use on-perm fast reasoning mode, handle dynamic job arrivals and real-time workload changes, validating performance on production traces, and exploring advanced constraint handling for job dependencies and energy-aware scheduling. 



\section*{Acknowledgments}
This work is supported by the U.S. Department of Energy, under grant \# DE-SC0024387.

\bibliographystyle{ACM-Reference-Format}
\bibliography{ref}




\end{document}